\documentclass[aps,showpacs,pra,twocolumn]{revtex4-1}  

\usepackage{amsmath}
\usepackage{mathrsfs}
\usepackage{amssymb}
\usepackage{graphicx,epstopdf,epsfig}
\usepackage{bm}
\usepackage{color}
\usepackage{times}
\usepackage{hyperref}
\usepackage{comment}

\newcommand{\bra}[1]{{\langle #1 \vert}} 		
\newcommand{\ket}[1]{{\vert #1 \rangle}} 		

\begin{document}
\title{Entanglement of neutral-atom qubits with long ground-Rydberg coherence times} 
\author{C. J. Picken}
\author{R. Legaie}
\author{K. McDonnell}
\author{J. D. Pritchard}
\address{SUPA, Department of Physics, University of Strathclyde, Glasgow G4 0NG, United Kingdom}

\date{\today}

\begin{abstract}
We report results of a ground-state entanglement protocol for a pair of Cs atoms separated by 6~$\mu$m, combining the Rydberg blockade mechanism with a two-photon Raman transition to prepare the $\ket{\Psi^+}=(\ket{10}+\ket{01})/\sqrt{2}$ Bell state with a loss-corrected fidelity of 0.81(5), equal to the best demonstrated fidelity for atoms trapped in optical tweezers but without the requirement for dynamically adjustable interatomic spacing. Qubit state coherence is also critical for quantum information applications, and we characterise both ground-state and ground-Rydberg dephasing rates using Ramsey spectroscopy. We demonstrate transverse dephasing times $T_2^*=10(1)$~ms and $T_2'=0.14(1)$~s for the qubit levels and achieve long ground-Rydberg coherence times of $T_2^*=17(2)~\mu$s as required for implementing high-fidelity multi-qubit gate sequences where a control atom remains in the Rydberg state while applying local operations on neighbouring target qubits.
\end{abstract}

\maketitle

\section{Introduction}
Neutral atoms are ideal candidates for quantum information processing, offering long coherence times as a quantum memory and ease of scaling to large numbers of qubits \cite{ladd10,weiss17}. Controllable long-range interactions can be engineered between the qubits by coupling to highly-excited Rydberg states which have extremely large dipole-moments, giving rise to a phenomenon known as dipole blockade~\cite{lukin01,urban09,gaetan09} whereby only a single Rydberg excitation can be created for atoms separated by $R\lesssim10~\mu$m. This mechanism provides an efficient route to implementing fast multi-qubit gate protocols~\cite{saffman10,saffman16} that are challenging in other platforms, enabling high efficiency realisation of Grover’s search algorithm \cite{molmer11,petrosyan16} and capable of achieving fault-tolerant computing using surface codes \cite{auger17}. These interactions have also been exploited to perform simulations of quantum magnetism \cite{labuhn16,bernien17,lienhard18,guardadosanchez18}. Rydberg atoms also offer strong coupling to superconducting microwave circuits enabling hybrid quantum computing \cite{petrosyan09,pritchard14} , efficient optical to microwave conversion \cite{gard17} and extended mm-scale interactions \cite{petrosyan08,sarkany15}.

Recently attention has focused on the implementation of single atoms trapped in fixed arrays and microscopic tweezer traps \cite{schlosser02} enabling arbitrary geometry arrays of variable size \cite{nogrette14,barredo14,barredo18} with deterministic assembly of defect-free arrays of over 50 qubits in 1D \cite{bernien17}, 2D \cite{barredo16} and 
 \cite{barredo18} as well as the ability to cool atoms to the vibrational ground-state \cite{kaufman12,thompson13,sompet17}. This platform is also compatible with dual-species operation \cite{zeng17} enabling controllable assembly of molecules \cite{liu18}.

Demonstration of ground-state entanglement in such systems has exploited Rydberg blockade \cite{wilk10,isenhower10,zhang10,maller15,zeng17} to achieve raw (corrected) fidelities of up to 0.73 (0.79) \cite{maller15} whilst Rydberg dressing has been used to obtain a post-selected fidelity of 0.81 \cite{jau16}. Despite their superior performance to entanglement based on local spin exchange \cite{kaufman15} or projective measurements \cite{lester18} using agile traps, the demonstrated Rydberg atom gates are limited by short coherence times $T_2^*<10~\mu$s \cite{johnson08} and laser induced dephasing \cite{leseleuc18} which affects experiments performing both gate-based computing and quantum simulation. {\color{black}{However, recent results show that suppression of this laser technical noise enables ground-Rydberg entanglement fidelities of up to 0.97 \cite{levine18} making Rydberg atoms a realistic candidate for scalable quantum computation.}}

In this paper we present results of a ground-state entanglement protocol for a pair of atoms separated by 6~$\mu$m, combining the Rydberg blockade mechanism with two-photon Raman transitions to prepare the $\ket{\Psi^+}=(\ket{10}+\ket{01})/\sqrt{2}$ Bell state with a loss-corrected fidelity of 0.81(5), equal to the best demonstrated fidelity for atoms trapped in optical tweezers but without the requirement of dynamically adjustable interatomic spacing \cite{jau16}. Qubit state coherence is also critical for quantum information applications, and we characterise both ground-state and ground-Rydberg dephasing rates using Ramsey spectroscopy, demonstrating qubit dephasing times of $T_2^*=10(2)$~ms and achieving extended ground-Rydberg coherence times of $T_2^*=17(2)~\mu$s compared to previous measurements \cite{johnson08} as required for implementing multi-qubit gate sequences where a control atom remains in the Rydberg state while applying local operations on neighbouring qubits \cite{saffman10,saffman16}. 

\begin{figure}[t!]
\centering
\includegraphics[width=8.6cm]{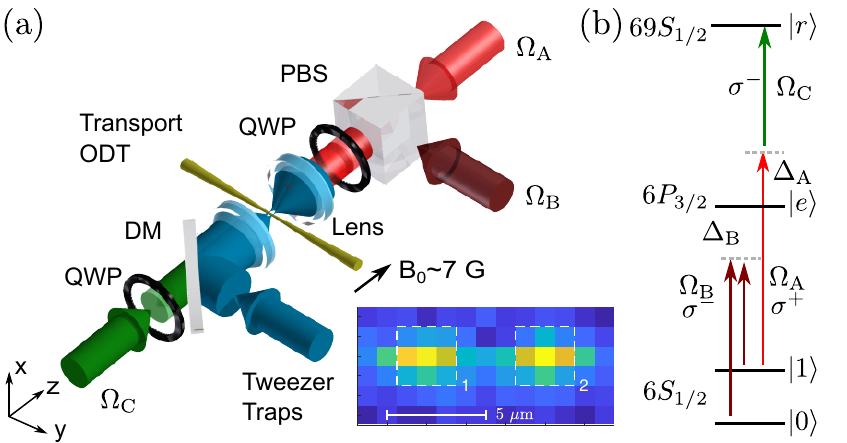}
\caption{(Color Online) (a) Experimental setup showing single atoms trapped in microscopic tweezer traps with 6~$\mu$m separation loaded from a large volume dipole trap. Excitation lasers are combined on a  polarising beam splitter (PBS) and counter propagate with a Rydberg excitation beam that is overlapped with trap light on a dichroic mirror (DM) and circularly polarized using quarter wave plates (QWP). (b) Qubit level scheme with information encoded in Cs hyperfine states $\ket{0}=\ket{F=3,m_F=0}$ and $\ket{1}=\ket{F=4,m_F=0}$. \label{fig:fig1}}
\end{figure}

\section{Experiment Setup}
The experiment setup is shown schematically in Figure~\ref{fig:fig1}(a), with single Cs atoms confined in a pair of orthogonally polarized microscopic tweezer traps at a wavelength of 1064~nm focused to a $1/e^2$ radius of 1.9~$\mu$m using high numerical aperture (NA) lenses and separated by $R=6~\mu$m. Atoms are initially cooled in a magneto-optical trap before being optically transported a distance of 30~cm to a low pressure science cell where they are loaded into the tweezer traps at a depth of $U_0=2$~mK. High-fidelity detection is achieved using fluorescence imaging to collect scattered light on an sCMOS camera as detailed in \cite{picken17}. After the initial readout, atoms undergo polarisation gradient cooling followed by an adiabatic lowering of the trap depth $U_0=300~\mu$K over 10~ms to reach temperatures of 5~$\mu$K. A bias field of $B_0=7.5$~G aligned along the $z$-axis is used to define the quantisation axis, ramping on in 10~ms followed by a 50~ms hold time to allow eddy currents to decay. Atoms are initialized in the $\ket{1}=\vert F=4,m_F=0 \rangle$ state via optical pumping with a single $\pi$-polarized beam resonant from $\ket{6S_{1/2},F=4}\rightarrow\ket{6P_{3/2},F'=4}$ propagating along the x-axis. We achieve a preparation fidelity of $\eta_\mathrm{OP}=95$~\%; however, as this beam is not retro-reflected atoms are heated to around 10~$\mu$K following state preparation. 

Qubit operations are performed using two-photon transitions via the $D_2$-line as shown in Figure~\ref{fig:fig1}(b). Ground-state rotations between the magnetically insensitive clock states $\vert 1 \rangle\rightarrow\vert 0 \rangle = \vert F=3,m_F=0 \rangle$ are driven using a Raman transition detuned by $\Delta_B/2\pi=-43$~GHz from the $6P_{3/2}$ transition. The Raman laser system at 852~nm uses an electro-optic modulator (EOM) to generate sidebands at 4.6~GHz before filtering the carrier using a {\color{black}{Mach-Zehnder}} interferometer \cite{haubrich00} to obtain co-propagating Raman beams with equal amplitude ($\Omega_B$). Rydberg excitation from $\ket{1}\rightarrow\ket{r}=\ket{69S_{1/2},m_j=1/2}$ uses light at 852~nm ($\Omega_A$) and 509~nm ($\Omega_C$) detuned by $\Delta_A/2\pi=870$~MHz from the $\ket{6S_{1/2},F=4}\rightarrow\ket{6P_{3/2},F'=5}$ transition to reduce spontaneous emission. The Rydberg excitation lasers are stabilized to a high finesse ULE cavity to obtain sub-kHz linewidths (See \cite{legaie18} for further details).

The excitation lasers are aligned through the same high NA lenses used to trap and image the atoms, with the 852~nm lasers for ground-state rotations ($\Omega_B$) and Rydberg excitation ($\Omega_A$) combined with orthogonal polarisation on a polarising beam splitter (PBS) and focused to $1/e^2$ waists of $20~\mu$m (15~$\mu$m) with $\sigma^-$($\sigma^+$) circular polarisation and powers of $200~\mu$W ($2~\mu$W) respectively. The 509~nm Rydberg excitation beam $\Omega_C$ is overlapped with the dipole trap beams using a dichroic mirror (DM) and focused to a $1/e^2$ waist of $18~\mu$m with a power of 90~mW, where the counter-propagating geometry minimizes Doppler sensitivity of the two-photon excitation. Laser intensity fluctuations are stabilized to around 1\% using active noise-eaters feeding back to acousto-optic modulators (AOMs), with additional AOMs to control pulse areas with a timing resolution of 25~ns.

Following initialisation in state $\ket{1}$, the dipole trap is turned off to eliminate differential AC Stark shifts and photoionisation or mechanical effects due to anti-trapping of the Rydberg state during which time rotation pulses are applied to the qubits. Due to the low single atom temperatures, we use drop times of up to $17.5~\mu$s whilst obtaining a recapture probability $P>0.95$. Atoms are recaptured at a depth of $U_0=1$~mK to provide efficient Rydberg state detection, using the anti-trapping potential to push any atoms in the Rydberg state out of the trapping volume on a timescale $t_\mathrm{recap}=8~\mu$s fast compared to the 300~K black-body limited lifetime $1/\Gamma_r=134~\mu$s \cite{beterov09} yielding a detection efficiency equal to $\eta_r=1-\Gamma_rt_\mathrm{recap}=94$~\% \cite{leseleuc18}. After recapture, the trap is again lowered to a depth of 300~$\mu$K to allow qubit state selection using a blow-away beam resonant with the $\ket{6S_{1/2},F=4}\rightarrow\ket{6P_{3/2},F=5}$ propagating along the $x$-axis to eject atoms in state $\ket{1}$ from the trap whilst leaving atoms in $\ket{0}$ unaffected. We verify the state selection efficiency $>99$\% for atoms in either state. All data points are extracted from 100-250 repeats of the experiment, with a single atom loading efficiency of 60\%.

\section{Coherent control of atomic qubits}

\begin{figure}
\centering
\includegraphics{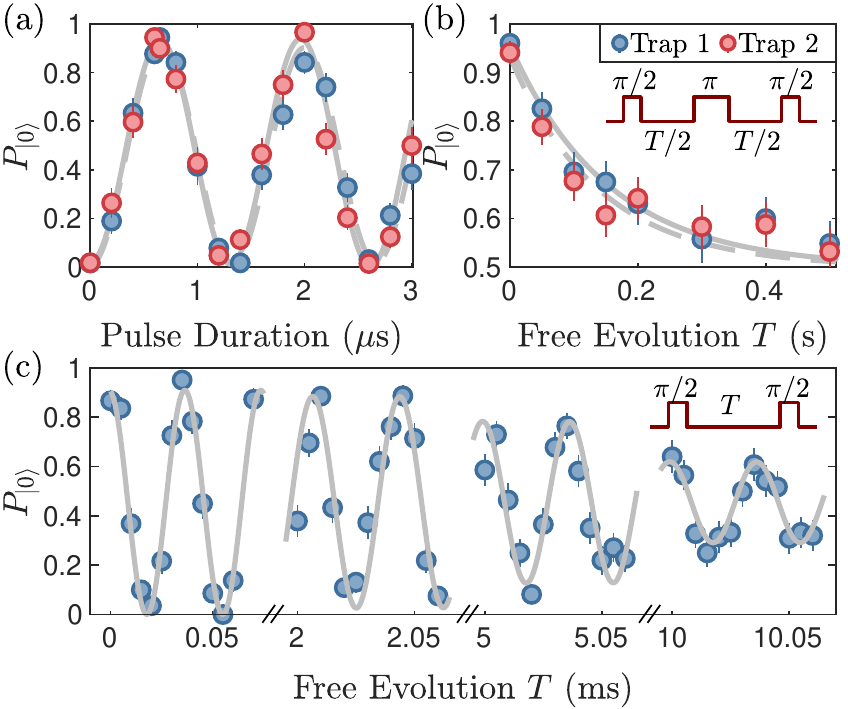}
\caption{(Color Online) (a) Ground-state rotations at $\Omega_{R}/2\pi=0.75(1),0.76(1)$~MHz for each trap (b) Ramsey measurement including spin-echo at $T/2$ yielding homogeneous dephasing time $T_2'=0.15(2),0.14(2)$~s from fitting an exponential decay (c) Ramsey sequence without spin-echo to obtain the inhomogeneous dephasing rate $T_2^*=10(1)~$ms for both traps. Insets show pulse sequence for $\Omega_B$. \label{fig:fig2}}
\end{figure}

A key requirement for quantum information is the ability to achieve long qubit coherence relative to the gate time. Here we demonstrate the application of the Raman laser to drive coherent rotations between $\ket{1}\rightarrow\ket{0}$ by measuring the population of state $\ket{0}$ as a function of the Raman pulse duration for both traps. Figure~\ref{fig:fig2}(a) shows the two traps undergoing synchronous oscillation with a two-photon Rabi frequency of $\Omega_{R}/2\pi=0.75(1),0.76(1)$~MHz for each trap, corresponding to a $\pi$-time of $t_\pi=660$~ns. The observed state transfer is 95\%, limited by the optical pumping efficiency. This limitation can be overcome by optically pumping on the $D_1$ line with larger excited state hyperfine splitting~\cite{zhang10}. 

Qubit coherence can be characterized in terms of longitudinal and transverse dephasing rates. The longitudinal decay rate $T_1$ is measured by calculating the probability for atoms in state $\ket{1}$ to reach state $\ket{0}$ due to off-resonant scattering from the dipole trap. We observe no spin-flips over the vacuum limited lifetime, placing a lower bound of $T_1>5$~s. To measure the transverse dephasing rates to extract the coherence time of the system we use Ramsey spectroscopy to apply two $\pi/2$-pulses separated by a time $T$, with (without) a spin-echo pulse of area $\pi$ at time $T/2$ to measure the homogeneous (inhomogeneous) dephasing rates $T_2'$ ($T_2^*$) \cite{kuhr05}. For these measurements, the atoms are held in the traps at a constant depth of 300~$\mu$K and data is recorded for both traps simultaneously. Figure~\ref{fig:fig2}(b) shows the probability to transfer into state $\ket{0}$ as a function of time for the case of a spin-echo Ramsey measurement. Fitting the data to an exponential decay yields a homogeneous dephasing time of $T_2'=0.15(2),0.14(2)$~s for each trap, limited by the residual differential trap shift which can be compensated using a `magic' magnetic field \cite{derevianko10,dudin10,yang16} or compensation beam \cite{kaplan02,carr16}. 

The inhomogeneous dephasing rate is extracted from the standard Ramsey sequence with no echo as shown in Figure~\ref{fig:fig2}(c), to which the fringe amplitude is fit using the equation \cite{kuhr05}
\begin{equation}
\alpha(T)=\left[1+0.95\left(\frac{T}{T_2^*}\right)^2\right]^{-3/2},\label{eq:T2star}
\end{equation}
resulting in $T_2^*=10(1)$~ms for both traps. This result is in excellent agreement with the predicted temperature-limited dephasing time equal to $T_2^*=0.97\times2\hbar/(\eta k_\mathrm{B}T)=10~$ms, where $\eta=1.45\times10^{-4}$ is the ratio of the qubit hyperfine splitting to the effective dipole trap detuning \cite{kuhr05}. Thus, even without an echo-pulse, the ratio of coherence time to gate time yields a ratio $>10^4$.

\begin{figure}[t]
\centering
\includegraphics{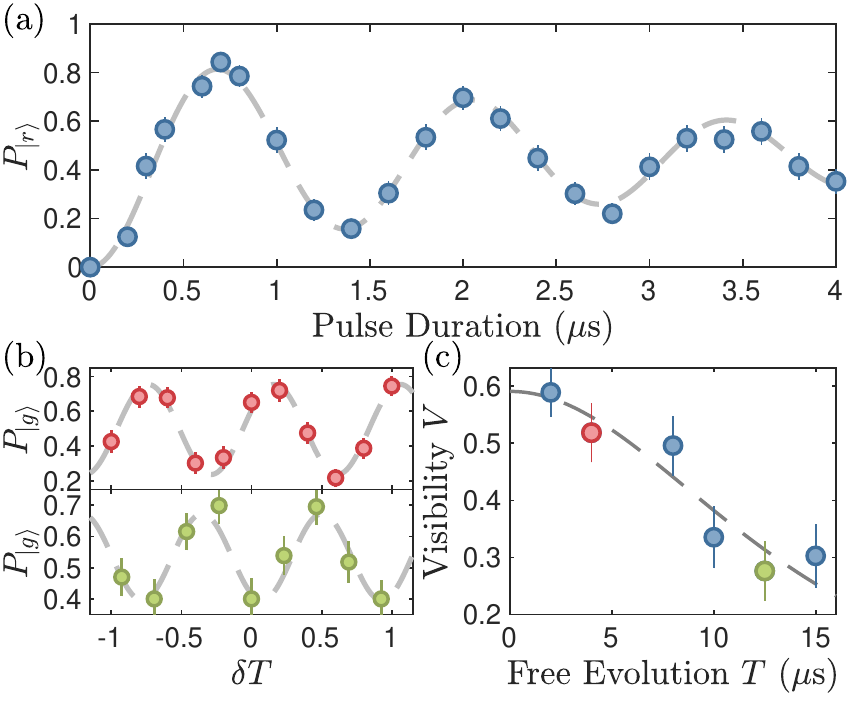}
\caption{(Color Online) Rydberg excitation (a) Single atom Rabi flop at $\Omega_{r}/2\pi=0.73(1)$~MHz with a $1/e$ damping rate of $\tau=3.2(3)~\mu$s (b) Ramsey fringes for $T=4~\mu$s (upper panel) and $T=12.5~\mu$s (lower panel) (c) Ramsey fringe visibility vs $T$ yielding a dephasing time $T_2^*=17(2)~\mu$s. \label{fig:fig3}}%
\end{figure}

A common limitation in experiments exploiting Rydberg atoms for quantum information is the finite transfer and detection efficiency of excitation to the Rydberg states, combined with fast dephasing of the Rydberg states. To characterize these effects in our system we first consider exciting a single atom to the Rydberg state $\ket{1}\rightarrow\ket{r}$, blocking trap 2 to avoid any effects due to long-range interactions with an atom in the other trap. Figure~\ref{fig:fig3}(a) shows oscillations in the $\ket{1}$ state as a function of the Rydberg excitation pulse duration, with an effective Rabi frequency $\Omega_\mathrm{r}/2\pi=0.73(1)$~MHz and a $1/e$ damping time of $\tau=3.2(3)~\mu$s. Assuming a 1D velocity distribution $\Delta v=\sqrt{k_\mathrm{B}T/m}\sim25~$mm/s and effective wavevector $k_\mathrm{eff}=k_{509}-k_{852}=5\times10^6~\mathrm{m}^{-1}$, this dephasing rate is significantly faster than that expected from the residual Doppler shift $k_\mathrm{eff}\Delta v\sim2\pi\times20$~kHz and independent of changes of the intermediate state detuning eliminating spontaneous emission. Following the analysis of Ref.~\cite{leseleuc18} this can be attributed to the residual phase-noise of the laser lock, verified by observing the dephasing rate to increase with increased feedback to the cavity mode. {\color{black}{This is verified by observing the damping rate to be insensitive to changes in intermediate state detuning (eliminating spontaneous emission) but decreasing with an increase in the electronic feedback applied to the laser diode.}} A consequence of this dephasing is to limit the maximum state transfer at $t_\pi$ to $P_\ket{r}=0.85$, rather than the peak of $\eta_\mathrm{OP}\eta_\mathrm{r}=0.89$ expected from finite state preparation and detection efficiency. This also limits the fraction of atoms returning to state $\ket{1}$ at $t_{2\pi}$. 

In addition to the laser induced dephasing rate, the Rydberg state coherence relative to $\ket{1}$ is critical to allow sequential gate operations \cite{saffman10}. To measure the dephasing we again use Ramsey spectroscopy to apply $\pi/2$ pulses separated by time $T$, however due to the fast phase-accumulation rate arising from the differential Stark-shift of the excitation lasers we record the fringes as a function of two-photon detuning $\delta$ for fixed $T$ as shown in Figure~\ref{fig:fig3}(b). For each time $T$ we extract the fringe visibility $V$, as shown in Figure~\ref{fig:fig3}(c). The visibility is fit using Eq.~\ref{eq:T2star} yielding a transverse dephasing time $T_2^*=17(2)~\mu$s, around twice the previous best reported Rydberg state coherence time \cite{johnson08}. This is limited by the residual thermal motion of the atom, with the measured coherence time comparable to the expected Doppler limited dephasing rate of $T_{2,D}=1/(\sqrt{2}\Delta v k_{\mathrm{eff}})\sim11~\mu$s~\cite{saffman11}. This residual dephasing places an upper-bound on the entangled state fidelity of $\mathcal{F}=[1+\exp(-t^2/T^2_2)]/2\gtrsim0.99$ for a $t\lesssim2~\mu$s pulse sequence as used below, showing we have attained dephasing rates suitable for high fidelity gates.

\begin{figure}[t]
\centering
\includegraphics{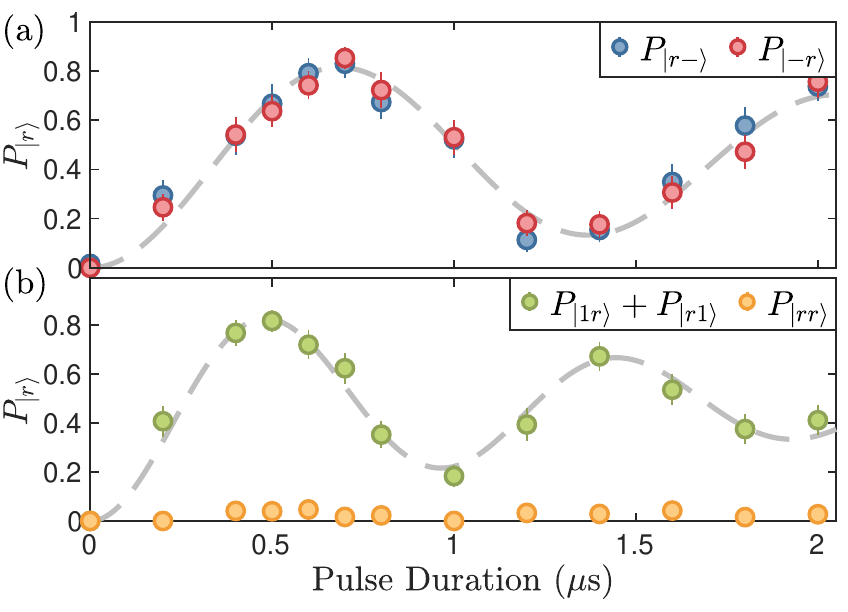}
\caption{(Color Online) Dipole Blockade (a) Single atom Rabi flop in each trap when only one atom is loaded. (b) Collective Rabi oscillation for Rydberg excitation of two atoms with $\Omega'_r/2\pi=1.04(2)$~MHz with suppression of excitation of $\ket{rr}$. \label{fig:fig4}}%
\end{figure}

\section{Entanglement via Rydberg Blockade}
Rydberg atoms experience strong long-range dipole-dipole interactions which in the Van der Waals regime scale as $V(R)=-C_6/R^6$, where $R$ is the interatomic separation and $C_6$ is the dispersion coefficient. In the limit $V(R)>\Omega_r$ these strong interactions give rise to dipole-blockade whereby only a single Rydberg state can be excited \cite{lukin01}, which corresponds to a blockade radius $R_\mathrm{b}=\sqrt[6]{\vert C_6\vert /\Omega_r}$. For two atoms excited simultaneously, the blockade mechanism causes oscillations between $\ket{11}$ and the symmetric entangled state $\ket{W}=(\ket{1r}+\ket{r1})/\sqrt{2}$ at a collectively enhanced rate of $\sqrt{2}\Omega_r$.

The interaction strength for atoms in the $\ket{69S_{1/2},m_j=1/2}$ state with an angle of $\pi/2$ between the quantisation axis and the internuclear axis is  $C_6=-573$~GHz$~\mu$m$^6$ \cite{sibalic17}, resulting in a blockade radius $R_\mathrm{b}=9.6~\mu$m. We demonstrate blockade for atoms at $R=6~\mu$m in Figure~\ref{fig:fig4}, where Figure~\ref{fig:fig4}(a) shows coherent Rabi oscillation in both traps at $\Omega_r$ when only a single atom is loaded, whilst for events where two atoms are initially loaded in the trap we see in Figure~\ref{fig:fig4}(b) an enhancement in the rate of transfer from $\ket{11}$ to $\ket{1r}$ and $\ket{r1}$ at $\Omega'_r/2\pi=1.04(2)$~MHz. Alongside this enhancement we observe a strong suppression of double excitation events with $P_{\ket{rr}}<5$\% for all times due to blockade. Taking the ratio of $\Omega_r'/\Omega_r$ yields $1.42(3)\sim\sqrt{2}$ as expected. We also measure the laser driven decay time for the collective state to be {\color{black}{approximately}} half of that measured for a single atom, giving $\tau=1.6(2)~\mu$s {{\color{black}{however numerical simulations show that this cannot be explained by spontaneous emission or differential Doppler shifts \cite{levine18}.}}}

\begin{figure}[t!]
\centering
\includegraphics{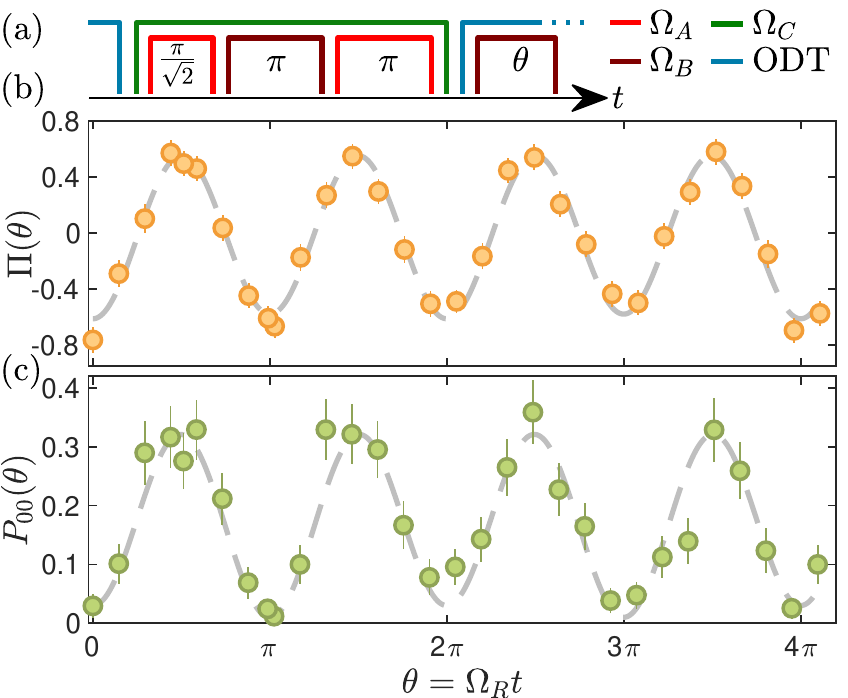}
\caption{(Color Online) Ground state entanglement. (a) Entanglement gate sequence (b) Parity oscillation showing evidence of off-diagonal coherence (c) Probability of observing atoms in $\ket{00}$. \label{fig:fig5}}%
\end{figure}

Measurement of the collective enhancement by itself is insufficient to demonstrate preparation of entanglement, and instead we use the pulse sequence shown in Figure~\ref{fig:fig5}(a) to map the $\ket{W}$ state onto a ground-state Bell state ${\ket{\Psi^+}=(\ket{01}+\ket{10})/\sqrt{2}}$. A Rydberg $\pi/\sqrt{2}$-pulse maps $\ket{11}\rightarrow\ket{W}$ followed by a $\pi$-pulse on the ground-state transition to populate $\ket{\psi}=(\ket{0r}+\ket{r0})/\sqrt{2}$, which is transformed to $\ket{\Psi^+}$ using a second Rydberg  $\pi$-pulse. The total sequence takes $1.85~\mu$s, during which time the coupling laser $\Omega_C$ remains on at constant amplitude and the Rydberg pulses are controlled using $\Omega_A$. Following the entanglement preparation sequence, atoms are recaptured and a global ground-state rotation of area $\theta=\Omega_R t$ is applied to the atoms. This enables us to observe oscillations in the parity, $\Pi=P_\ket{00}+P_\ket{11}-P_\ket{10}-P_\ket{01}$, to extract the off-diagonal coherences of the density matrix $\rho$ required to determine the fidelity \cite{turchette98}. Figure~\ref{fig:fig5}(b) shows the resulting parity oscillation fit to $\Pi(\theta)=P_0+A\cos(\Omega_Rt)+B\cos(2\Omega_Rt)$ where for a perfect $\ket{\Psi^+}$ state we expect $P_0=A=0$ and $B=-1$. We observe a strong oscillation at $2\Omega_R$; however, the contrast is limited due to finite state preparation and losses.

\begin{table}[b] 
\begin{ruledtabular}
\begin{tabular}{l c}
Matrix Elements & Measured Value\\
\hline
$\rho_{00,00}=P_\ket{00}=P_{00}(0)$& 0.03(2) \\
$\rho_{11,11}=P_\ket{11}=P_{00}(\pi)$ &   0.01(1) \\
$\rho_{01,01}+\rho_{10,10}=P_\ket{01}+P_\ket{10}$ &0.72(3)\\
$\mathcal{R}(\rho_{01,10})$ & 0.27(3)\\
$\mathcal{F}=(P_\ket{01}+P_\ket{10})/2+\mathcal{R}(\rho_{01,10})$&0.63(3)\\
$\mathcal{F}_{pairs}=\mathcal{F}/p_\mathrm{recap}$ & 0.81(5)\\
\end{tabular}
\caption{Measured values of the density matrix extracted from analysis of the parity oscillation data of Fig.~\ref{fig:fig5}\label{tab1}}.
\end{ruledtabular}
\end{table}

The Bell-state fidelity is given by $\mathcal{F}=\bra{\Psi^+}\rho\ket{\Psi^+}=(P_\ket{01}+P_\ket{10})/2+\mathcal{R}(\rho_{01,10})$ where $\mathcal{R}(\rho_{01,10})$ is the real-part of the off-diagonal coherence. A simple analysis shows that this coherence can be extracted directly from the amplitude parity oscillation giving $\mathcal{F}\sim0.76$ \cite{turchette98,maller15}. However, in the presence of loss the readout method of counting atoms following the blow-away returns false positives leading to over-estimation of population of either atom in state $\ket{1}$. We therefore adopt the analysis method of \cite{gaetan10} to determine the fidelity in the presence of loss. Loss rates $L_{1,2}=0.13(1),0.12(1)$ for each trap are calculated from the mean probability of finding atoms in either trap averaged over $\theta$, resulting in total loss $L_t=1-\mathrm{tr}(\rho)=L_1+L_2-L_1L_2=0.24(1)$. This agrees well with the recapture probability for a pair of atoms in the absence of the state-selective blow-away pulse, $p_\mathrm{recap}=0.78(3)$. We now focus analysis on the $P_{00}(\theta)$ data shown in Figure~\ref{fig:fig5}(c) corresponding to both atoms being present and thus no loss occurring. Taking $P_\ket{00}=P_{00}(0)=0.03(2)$ and $P_\ket{11}=P_{00}(\pi)=0.01(1)$ we obtain $P_\ket{01}+P_\ket{10}=1-P_\ket{11}-P_\ket{00}-L_t=0.72(3)$. The mean value $\langle P_{00}(\theta)\rangle=(P_\ket{01}+P_\ket{10}+3(P_\ket{11}+P_\ket{00})+2\mathcal{R}(\rho_{01,10}))/8$ from which we can extract $\mathcal{R}(\rho_{01,10})=0.27(9)$. This can be verified independently using a similar approach to fit the parity oscillation data of Figure~\ref{fig:fig5}(b) using measured loss rates which yields $\mathcal{R}(\rho_{01,10})=0.27(3)$, resulting in a fidelity $\mathcal{F}=0.63(3)$, in excess of the bound $\mathcal{F}>0.5$ required for entanglement. Correcting for the loss, we obtain $\mathcal{F}_\mathrm{pairs}=\mathcal{F}/p_\mathrm{recap}=0.81(5)$, the highest post-selected ground-state entanglement generated using the dipole blockade mechanism and equal to that obtained using Rydberg dressing \cite{jau16}. These results are summarized in Table~\ref{tab1}.

The dominant error in the Bell state preparation arises from imperfect Rydberg excitation due to technical noise in the laser which limits the fraction of atoms returning to the $\ket{1}$ state following the second Rydberg $\pi$-pulse. This is also the cause of atom loss in the sequence, applying the pulse sequence with $\Omega_C$ off we retain atom pairs $p_\mathrm{recap}>99\%$ {\color{black}{showing no loss due to off-resonant scattering}}, whilst all operations are completed on a timescale fast compared to both ground and Rydberg state dephasing times. 

\section{Conclusion}
In conclusion, we have demonstrated ground-state entanglement of two atoms using the Rydberg blockade mechanism in a system achieving long coherence times in both the ground-state and Rydberg state. We obtain a Doppler limited ground-Rydberg dephasing rate of $T_2^*=17(2)~\mu$s, sufficiently long to implement high fidelity gates as required for quantum information processing. Further improvements in coherence time will require cooling the atoms to the motional ground-state prior to initialisation \cite{kaufman12,thompson13,sompet17}. Our entanglement sequence prepares 72\% of atom pairs into the $\ket{\Psi^+}$ state with a fidelity of $0.81(5)$, equal to the best demonstrated fidelity for atoms trapped in optical tweezers and currently limited by technical phase noise on the excitation lasers.

Whilst preparing this manuscript we became aware of recent work demonstrating suppression of technical noise in the Rydberg excitation lasers using light filtered by a high-finesse cavity to extend the damping time arising from laser induced dephasing to $27~\mu$s and demonstrating $\ket{W}$-state preparation fidelity of 0.97 \cite{levine18}. In future we plan to upgrade our laser system to improve the coherence of the driven ground-Rydberg transition and couple the Rydberg atoms to a superconducting microwave cavity, where the extended coherence times achieved in our experiment are well suited to implementing the weaker cavity mediated entanglement protocols enabling entanglement over length scales much larger than the blockade distance \cite{petrosyan08,sarkany15}.

\section{Acknowledgements}
This work is supported by the UK Engineering and Physical Sciences Research Council, Grant No. EP/N003527/1. The authors would like to thank Paul Griffin and Andrew Daley for useful comments. The data presented in the paper are available for download at \href{http://dx.doi.org/10.15129/936e9230-847a-4089-a17e-f7c625e1ef10}{http://dx.doi.org/10.15129/936e9230-847a-4089-a17e-f7c625e1ef10}.

%

\end{document}